# A Comparative Analysis of Transformer-less Inverter Topologies for Grid-Connected PV Systems: Minimizing Leakage Current and THD

Shashwot Shrestha[1*], Rachana Subedi[2], Swodesh Sharma[3], Sushil Phuyal[4], Indraman Tamrakar[5]

[1]Dept of Electrical Engineering, Pulchowk Campus, Tribhuvan University, Email: shashwotshrestha.7@gmail.com
[2]Dept of Electrical Engineering, Pulchowk Campus, Tribhuvan University, Email: rachana.subedi.29@gmail.com
[3]Dept of Electrical Engineering, Pulchowk Campus, Tribhuvan University, Email: xarmaswodesh@gmail.com
[4]Dept of Electrical Engineering, Pulchowk Campus, Tribhuvan University, Email: sushilphuyal.sp@gmail.com
[5]Professor, Dept. of Electrical Engineering, Pulchowk Campus, IOE, TU. Email: imtamrakar@ioe.edu.np

*Abstract— The integration of distributed energy resources (DERs), particularly photovoltaic (PV) systems, into the power grids have gained major attention due to their environmental and economic benefits. Although traditional transformer-based grid-connected PV inverter provides galvanic isolation for leakage current, it suffers from major drawbacks of high cost, lower efficiency, and increased size. Transformer-less grid-connected PV inverter (TLGI) has emerged as a prominent alternative as this achieves higher efficiency, compact design, and lower cost. However, due to a lack of galvanic isolation, TLGI is highly affected by the leakage current caused by the fluctuation of common mode voltage (CMV). This paper investigates three topologies —H4, H5, and HERIC— with the comparisons between their CMV, differential mode voltage (DMV), total harmonic distortion (THD), and leakage current. A simulation was done for each topology in MATLAB(Simulink) R2023a and the results from the simulation demonstrate that the H5 topology achieves a balance between low leakage current, reduced THD, and optimal operational efficiency, making it suitable for practical application.*

I. INTRODUCTION

Distributed energy resources (DERs), such as solar and wind energy systems, have become essential in transitioning to sustainable energy solutions. As global energy demands increase and concerns over climate change intensify, the need for clean, renewable, and sustainable energy sources has never been greater [1]. DERs are pivotal in decentralizing energy generation and reducing reliance on traditional fossil fuel-based power systems associated with high greenhouse gas emissions and environmental degradation [2].

Among DERs, photovoltaic power generation (PVPG) systems stand out due to their numerous advantages. Photovoltaic (PV) systems harness solar energy, a virtually inexhaustible and clean energy source, to generate electricity. These systems are increasingly being deployed in residential and commercial settings because of their ability to integrate seamlessly with existing infrastructure and their scalability. Whether installed on rooftops of homes or in large-scale solar farms, PV systems provide a reliable and sustainable solution to meet the growing energy demands of modern society [3]. Due to the advantages of PV, grid-connected PV systems are used, which inject energy into the power grids. To inject solar energy into the grid, firstly the DC voltage generated by the PV is converted to AC with the help of an inverter. After transferring to AC voltage, this voltage gets synchronized with the grid voltage so that the PV system can efficiently inject the energy into the grid [4].

Traditionally, PV inverters incorporate transformers to achieve galvanic isolation. This isolation is crucial as it separates the DC side (PV array) from the AC side (grid), thereby enhancing safety, minimizing leakage currents, and reducing electromagnetic interference (EMI) [5]. Transformer-based inverters, however, come with significant drawbacks. They are bulky, heavy, and costly, and their efficiency is often lower due to the energy losses associated with the transformer itself. These limitations have spurred the development of transformer-less inverters (TLIs), which eliminate the need for a transformer and offer several advantages.

Transformer-less inverters have revolutionized PV technology by addressing many of the shortcomings of traditional transformer-based designs. By removing the transformer, TLIs achieve a smaller footprint, reduced weight, and lower manufacturing costs. These benefits translate to easier installation, reduced transportation expenses, and improved overall system efficiency. Additionally, TLIs enable higher power densities and are better suited for integration into modern, compact PV systems [6].

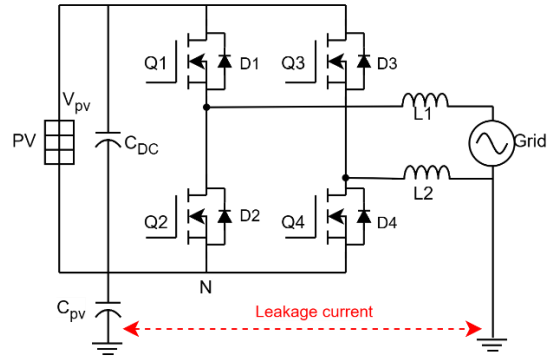

Fig. 1 Flow of leakage current in H4 inverter

Transformer-less inverters however have a major disadvantage regarding the leakage current. Transformer-less inverters result in a galvanic connection between the PV and grid side which leads to the formation of a closed loop between the grid and PV side through which the leakage current can flow. This flow of leakage current is highly affected by the parasitic capacitance, which forms between the PV panel and the ground. Due to the high-frequency switching of the inverter, the voltage between the PV panel and the ground fluctuates, which causes the leakage current to flow through the parasitic capacitance [7]. Fig. 1 shows the flow of leakage current in a conventional H4 topology grid-connected transformer-less inverter.

The emergence of leakage current can also be viewed from the point of view of Common Mode Voltage (CMV), which is the average potential difference between the inverter's terminals and ground. Leakage current is produced by the coupling of fluctuating CMV with the parasitic capacitance. CMV fluctuations result from the switching actions of the inverter's power electronics, and these fluctuations create a path for leakage current to flow through the parasitic capacitance. Hence, the leakage current can be reduced significantly if the CMV is maintained constant [8]. Another important factor determining the efficiency of a transformer-less inverter is its output voltage also called differential mode voltage (DMV). The nature of the waveform of DMV determines the total harmonic distortion (THD) generated by the inverter.

Leakage currents in PV systems can have several adverse effects. They contribute to efficiency losses by dissipating energy that could otherwise be used for power generation. Additionally, leakage currents pose significant safety hazards, as they can lead to electric shocks and compromise the stability of the system. Furthermore, high leakage currents can increase electromagnetic interference (EMI), which may disrupt nearby electronic devices and communication systems. From a regulatory standpoint, leakage currents must be controlled to comply with international safety standards, such as VDE 0126-1-1, which limits leakage currents to 300 mA [9].

Leakage current can be reduced by implementing various types of inverter topologies. In all these topologies the switching pattern of the switches of the inverter plays a crucial role in determining the nature of CMV and hence leakage current. The paper [10] presents the LC resonant circuit modeling of a traditional H4 bridge inverter to demonstrate the mathematical relationship between the leakage current and the CMV. The modeling involves the conversion of voltage sources into current sources and vice versa to develop the relationship between the CMV and the leakage current. Fig. 2 illustrates the simplified resonant circuit of the H4 inverter along with the mathematical expression of the derived leakage current

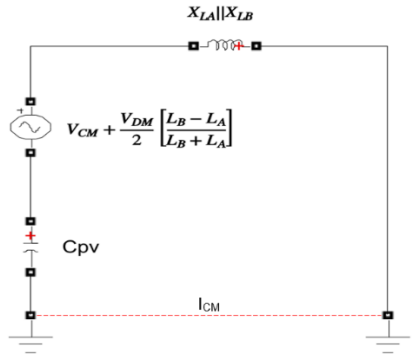

Fig. 2 Flow of leakage current in H4 inverter

$$I_{CM} = \frac{V_{tCM}}{(X_{LA} \| X_{LB}) + X_{C_{PV}}} \ldots\ldots\ldots(i)$$

$$V_{tCM} = V_{CM} + \frac{V_{DM}}{2}\left(\frac{L_B - L_A}{L_B + L_A}\right) \ldots\ldots\ldots(ii)$$

From equation (i), leakage current ($I_{CM}$) depends upon the magnitude as well as the nature of CMV ($V_{CM}$). When the CMV is almost constant in each mode of operation in both positive and negative half cycles, then it will be a DC quantity with a resonant frequency being zero. Therefore, the whole of the denominator of equation (i) becomes infinite, and hence leakage current tends to zero.

This paper discusses three topologies named (i) H4 topology, (ii) H5 topology, and (iii) HERIC topology. Also, their advantages and disadvantages in terms of total harmonic distortion (THD) are discussed by comparing the results obtained from the MATLAB (Simulink) simulation.

## II. MATERIALS AND METHODS

sAmong the various inverter topologies, not all of them can maintain the CMV constant. Since the elimination of leakage current requires constant CMV, leakage current exists in some inverter topology. In this section, three different inverter topologies and their CMV, DMV, THD, and RMS values of leakage current are compared.

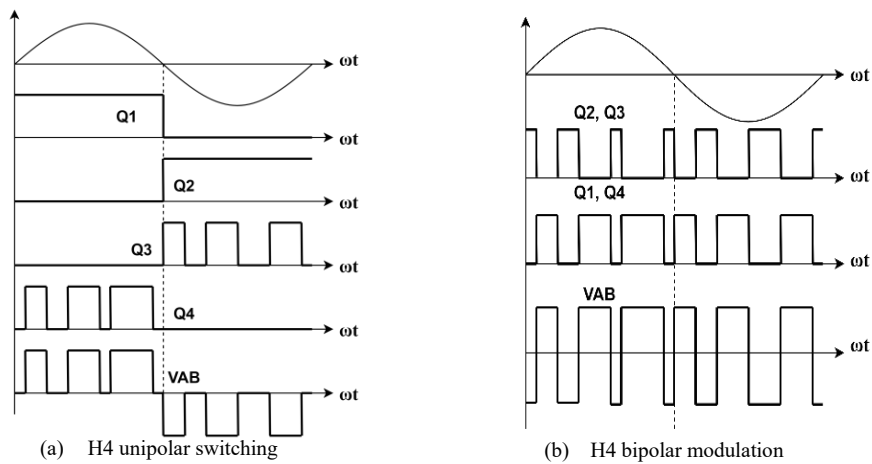

(a) H4 unipolar switching  (b) H4 bipolar modulation

Fig. 3 Switching pattern in H4 topology

## A. Topologies

1) **H4 Topology**: Fig. 1 shows the circuit diagram of H4 topology. It is one of the conventional topologies used for grid-connected inverters. It consists of four switches arranged in a full-bridge configuration. This topology consists of two modulation techniques, which are unipolar modulation and bipolar modulation. Fig. 3 shows the switching pattern of unipolar and bipolar modulation [11].

In unipolar modulation, there are four modes of operation. The four modes of operation are illustrated in Fig. 4. Mode 1 and 3 are energy delivery stages. At these stages, the energy is fed forward from the DC side through the inverter circuit to the grid. Modes 2 and 4 are the freewheeling stages, where the energy stored on the inductor is released to the grid. It is worth noting that the switches Q1 and Q2 operate at line-frequency while the switches Q3 and Q4 operate at high frequency as shown by Fig. 3a.

During mode 1, switches Q1 and Q4 are turned on while switches Q3 and Q2 are turned off. Fig. 4 shows the direction of flow of the inverter current $I_{inv}$. The CMV during this mode is given by the equation (i).

$$CMV = \frac{V_{AN}+V_{BN}}{2} \quad \ldots\ldots(iii)$$

$$CMV = \frac{V_{pv}+0}{2} = \frac{V_{pv}}{2}$$

The DMV during this mode is given by equation (ii).

$$DMV = V_{AN} + V_{BN} \quad \ldots\ldots(iv)$$

$$DMV = V_{PV}$$

The inverter output voltage or Differential Mode Voltage (DMV) is given by the difference of pole voltages $V_{AN}$ and $V_{BN}$.

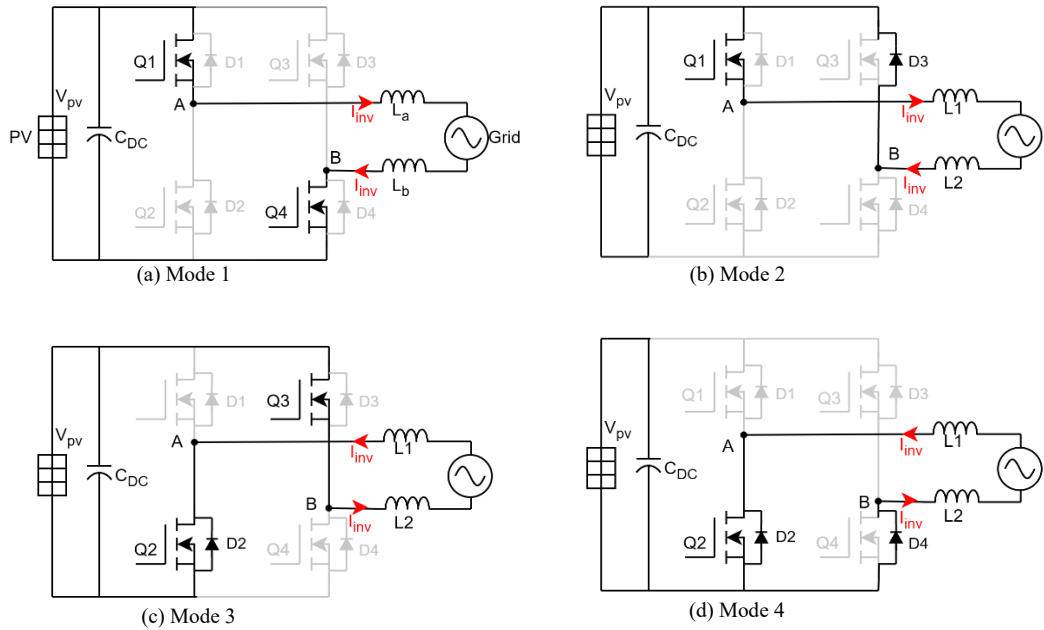

Fig. 4 Different Modes of Operation in H4 Unipolar topology

In mode 2, all the switches except Q1 are closed. The current supplied by the inductor flows through the switch Q1 and the freewheeling diode D3 of switch Q3. At this instant, the output voltage of the inverter (DMV) is zero and the CMV is given by:

$$CMV = \frac{V_{AN} + V_{BN}}{2} = V_{PV}$$
$$DMV = V_{AN} - V_{BN} = 0$$

In mode 3, energy is transferred from the PV side to the grid side through the combination of switches Q2 and Q3. At this instance, the output voltage of the inverter (DMV) is $-V_{PV}$, and CMV at this mode is given by:

$$CMV = \frac{0 + V_{PV}}{2} = \frac{V_{PV}}{2}$$
$$DMV = V_{AN} - V_{BN} = -V_{PV}$$

And in mode 4, both the pole voltages are zero. So, CMV = DMV = 0

The voltage characteristics of the H4 topology inverter incorporating unipolar modulation are summarized in Table 1.

Table 1: Voltage characteristics of H4 unipolar modulation

| Modes | $V_{AN}$ | $V_{BN}$ | DMV | CMV |
|---|---|---|---|---|
| 1 | $V_{PV}$ | 0 | $V_{PV}$ | $V_{PV}/2$ |
| 2 | $V_{PV}$ | $V_{PV}$ | 0 | $V_{PV}$ |
| 3 | 0 | $V_{PV}$ | $-V_{PV}$ | $V_{PV}/2$ |
| 4 | 0 | 0 | 0 | 0 |

From Table 1, it is evident that the CMV is not constant in this modulation. So, there is a significant amount of leakage current. From the DMV column, it is evident that the voltage takes three distinct levels in one complete cycle. This is desirable since it reduces the electrical stress on the inverter switches and leads to higher conversion efficiency [12]. Thus, unipolar modulation has desirable DMV characteristics and undesirable CMV characteristics.

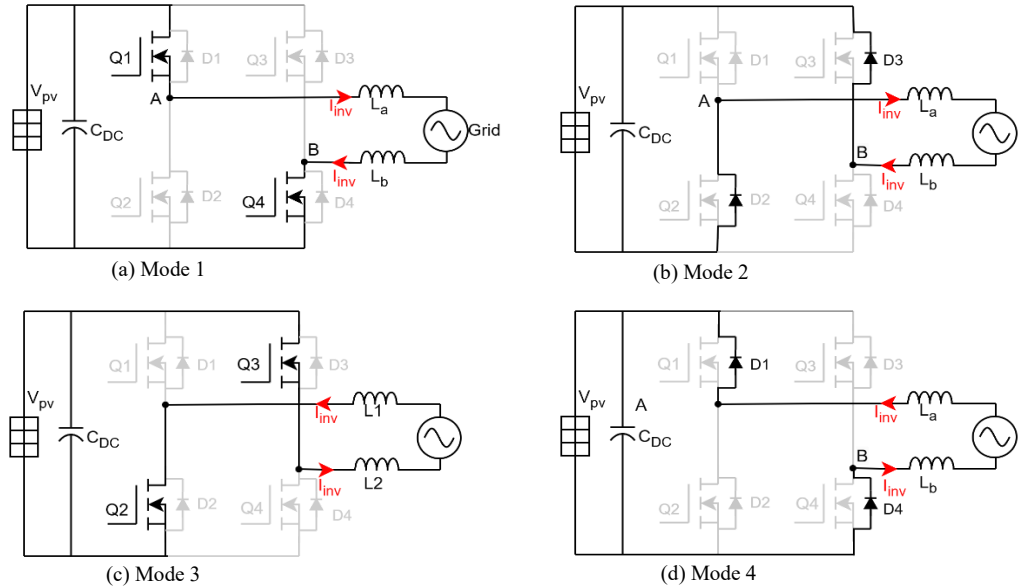

(a) Mode 1  (b) Mode 2  (c) Mode 3  (d) Mode 4

Fig. 5 Different Modes of Operation in H4 Bipolar topology

Fig. 3b depicts the switching pattern and DMV of bipolar modulation. Similar to unipolar modulation, the four modes of operation are shown in Fig. 5.

Similar to the unipolar modulation, modes 1 and 3 are energy delivery stages, at which the energy is fed forward by the DC side through the inverter circuit to the grid. Modes 2 and 4 are the energy return stages in which the energy is fed back into the DC side from the AC filter. Here, all switches operate at high frequency.

The voltage characteristics of H4 topology inverter incorporating bipolar modulation is summarized in Table 2.

Table 2: Voltage characteristics of H4 bipolar modulation

| Modes | $V_{AN}$ | $V_{BN}$ | DMV | CMV |
|---|---|---|---|---|
| 1 | $V_{PV}$ | 0 | $V_{PV}$ | $V_{PV}/2$ |
| 2 | 0 | $V_{PV}$ | $-V_{PV}$ | $V_{PV}/2$ |
| 3 | 0 | $V_{PV}$ | $-V_{PV}$ | $V_{PV}/2$ |
| 4 | $V_{PV}$ | 0 | $V_{PV}$ | $V_{PV}/2$ |

The CMV characteristics from Table 2 shows that bipolar modulation achieves constant CMV thereby ensuring low leakage current. However, the DMV has a two-level voltage waveform at switching frequency scale. This leads to generation of higher-level harmonics in the output voltage and the energy returning stages leads to higher reactive power swapping [13]. Additionally, the high switching frequency of the switches increases the switching loss, and the two-level voltage applies higher switching stress to the switches.

In the H4 topology, the CMV characteristics of bipolar modulation are favorable, since in this modulation the leakage current is almost zero due to constant CMV. However, due to poor DMV characteristics, this technique cannot be implemented in transformer-less grid-connected inverters since the output voltage have higher harmonic contents and the reactive power flow between the grid and PV-side is high along with higher switching losses. So, the goal in transformer-less grid-connected PV inverters is to suitably incorporate CMV characteristics of bipolar modulation and DMV characteristics of unipolar modulation. This technique can be achieved by using two new topologies "H5" and "HERIC" topologies [14].

2) *H5 Topology*: Inverter with this topology uses five switches contrary to H4 topology, where only four switches are used. This additional switch Q5 is inserted at the positive DC bus as shown in Fig. 6. This

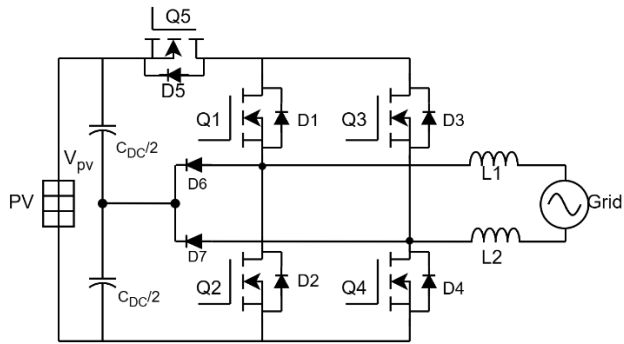

Fig. 6 H5 topology

switch helps to decouple the grid and PV side during the freewheeling period and weakens the CMV

source, which helps to reduce the leakage current [15]. Additionally, two clamping diodes ($D_6$ and $D_7$) are added as shown in Fig. 6. The function of these diodes is to clamp the pole voltages $V_{AN}$ and $V_{BN}$ to $\frac{V_{PV}}{2}$ during the freewheeling period, hence, to keep the CMV at a constant level during entire modes of operation [13]. The upper diode D6 clamps the upper pole voltage $V_{AN}$, while the lower diode D7 clamps the lower pole voltage $V_{BN}$ [16].

The modulation of H5 topology is such that it inherits the DMV characteristics of unipolar modulation and CMV characteristics of bipolar modulation. The waveforms of DMV and switching patterns are shown in Fig. 7. This topology also has four modes of operation, which are illustrated in Fig. 8.

During mode 1, the energy is transferred from the PV side to the grid side through the combination of switches Q5, Q1, and Q4. The output voltage of the inverter (DMV) is equal to $V_{pv}$ at this instant and the CMV $= \frac{V_{AN} + V_{BN}}{2} = \frac{V_{PV} + 0}{2} = \frac{V_{PV}}{2}$.

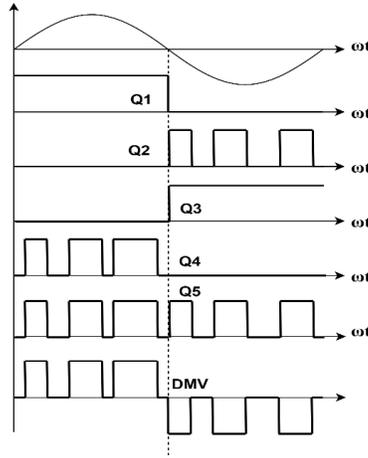

Fig. 7 Switching Pattern in H5 topology

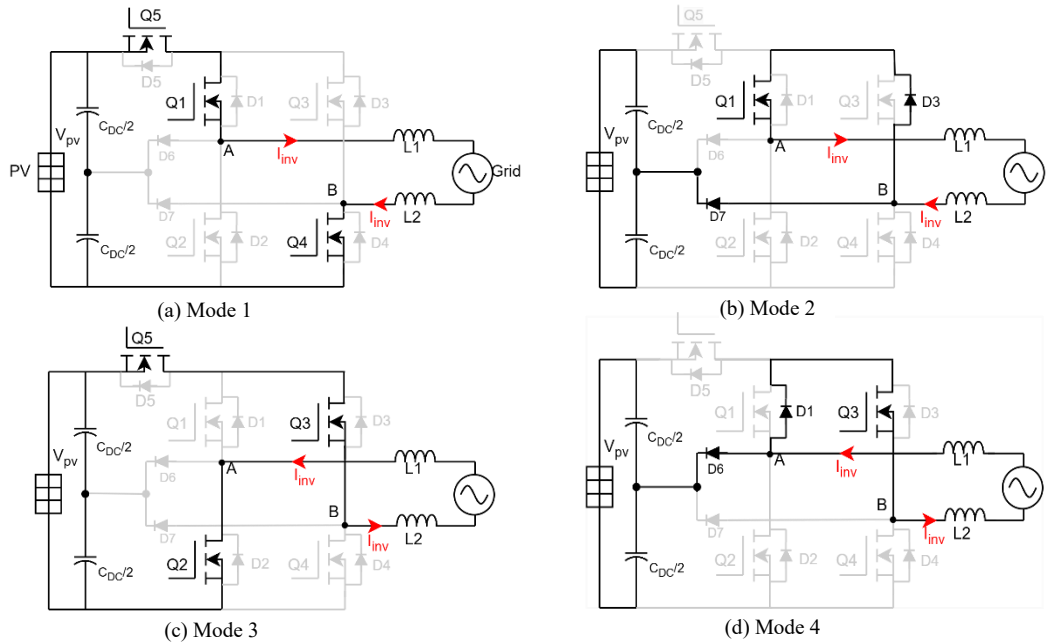

Fig. 8 Different Modes of Operation in H5 topology

In mode 2, among all the switches only Q1 conducts. This is the freewheeling period and the current supplied by the inductor flows through the switch Q1, diode D3, and the clamping diode D7. Since the switch Q5 is off, the DC side is decoupled from the AC side. The lower clamping diode D7 clamps the pole voltage $V_{BN}$ to $\frac{V_{PV}}{2}$. The output voltage VDM is zero at this instant and the CMV is $\frac{\frac{V_{PV}}{2} + \frac{V_{PV}}{2}}{2} = \frac{V_{PV}}{2}$.

In mode 3, the energy is transferred from the PV side to the grid side through the combination of switches Q5, Q2, and Q3. The output voltage DMV is equal to -$V_{PV}$, and CMV = $\frac{0 + V_{PV}}{2} = \frac{V_{PV}}{2}$.

Finally in freewheeling mode 4, only switch Q3 conducts, and the freewheeling current flows through switch Q3, diode D1, and the clamping diode D6. Also, in this mode, the AC and DC sides are decoupled. Here, DMV equals zero and CMV is $\frac{V_{PV}}{2}$. In this modulation only switches Q1 and Q3 operate at line-frequency, while the other operates at high frequency.

Table 3: Voltage characteristics of H5 modulation

| MODES | $V_{AN}$ | $V_{BN}$ | DMV | CMV |
|---|---|---|---|---|
| 1 | $V_{PV}$ | 0 | $V_{PV}$ | $V_{PV}/2$ |
| 2 | $V_{PV}/2$ | $V_{PV}/2$ | 0 | $V_{PV}/2$ |
| 3 | 0 | $V_{PV}$ | -$V_{PV}$ | $V_{PV}/2$ |
| 4 | $V_{PV}/2$ | $V_{PV}/2$ | 0 | $V_{PV}/2$ |

The voltage characteristics of H5 topology in summarized in Table 3. It can be seen that the DMV maintains three level voltage, which is desirable from the point of view of total harmonic distortion (THD). The CMV has constant values in all modes. Hence the leakage current in this topology is minimum [17].

Hence, H5 topology is able to maintain three level DMV (i.e., unipolar characteristics) and constant CMV (i.e., bipolar characteristics). Hence the THD and leakage current in this topology is minimum.

3) **HERIC Topology**: Here, six switches are used. The additional two switches Q5 and Q6 are used for decoupling the AC and DC sides. These switches are connected in an anti-series fashion. The decoupling strategy used in this topology is called "AC decoupling type". Thus, unlike the H5 topology which uses DC decoupling, this topology uses AC decoupling to weaken the CMV source during the freewheeling period. Fig. 9 depicts the circuit diagram of HERIC topology.

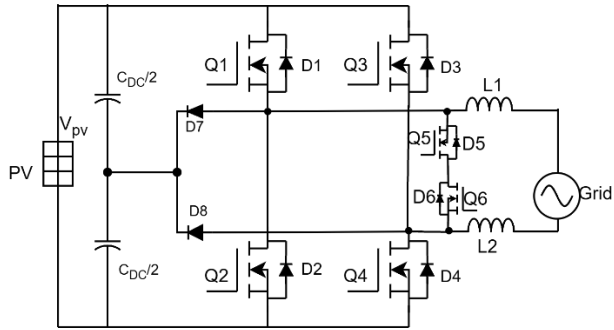

Fig. 9 Different Modes of Operation in HERIC topology

Like H5 topology, here two clamping diodes (D7 and D8) are used to clamp the CMV to $V_{pv}/2$ during the freewheeling period. The switches Q1-Q4 operate at high frequency while the two added switches

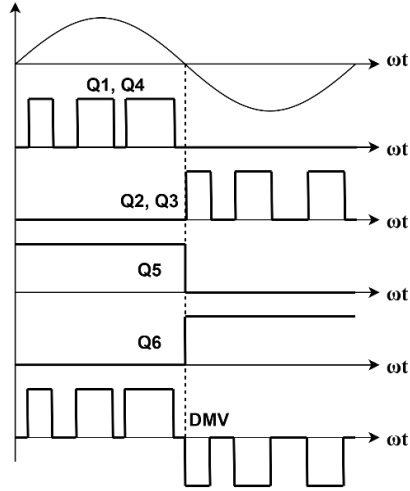

Fig. 10 Switching Pattern in HERIC topology

Q5 and Q6 operate at line-frequency. The waveforms of DMV and switching patterns are shown in Fig. 10 and four modes of operation are illustrated in Fig. 11. Also, similar to the H5 topology, modes 1 and 3 are energy delivering states and modes 2 and 4 are freewheeling stages.

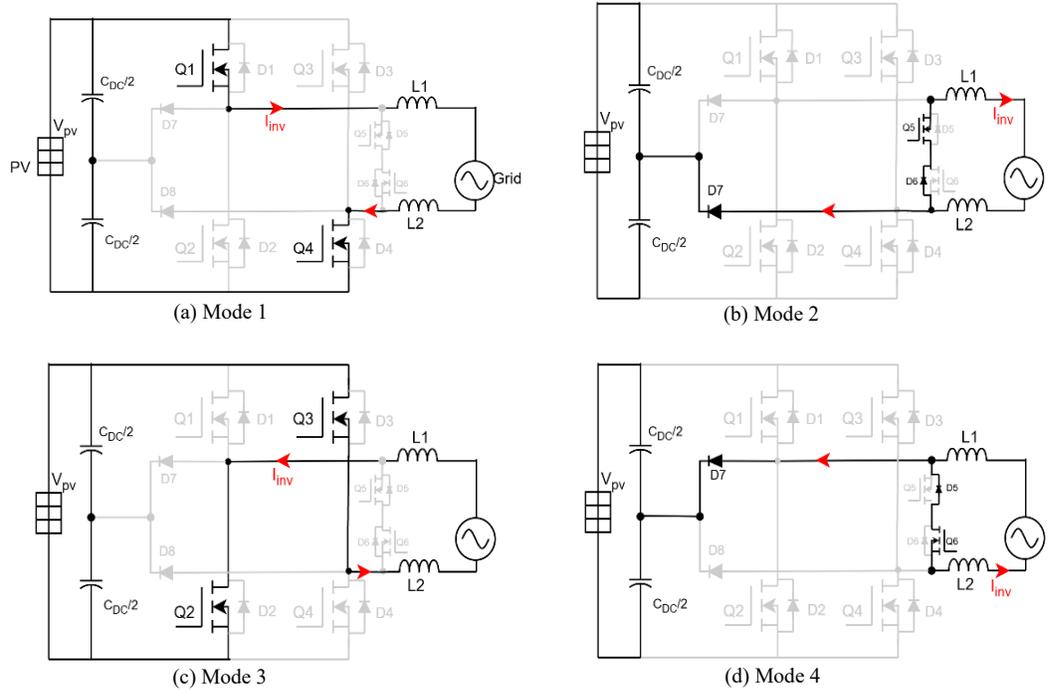

Fig. 11 Different modes of operation in HERIC topology

Voltage characteristics are tabulated in Table 4 for this mode.

Table 4: Voltage characteristics of HERIC topology

| Modes | $V_{AN}$ | $V_{BN}$ | DMV | CMV |
|---|---|---|---|---|
| 1 | $V_{PV}$ | 0 | $V_{PV}$ | $V_{PV}/2$ |
| 2 | $V_{PV}/2$ | $V_{PV}/2$ | 0 | $V_{PV}/2$ |
| 3 | 0 | $V_{PV}$ | $-V_{PV}$ | $V_{PV}/2$ |
| 4 | $V_{PV}/2$ | $V_{PV}/2$ | 0 | $V_{PV}/2$ |

It is evident that the CMV and DMV characteristics of this mode exactly matches with that of H5 topology. So, like H5 topology, HERIC topology is also able to maintain three levels of DMV (i.e., unipolar characteristics) and constant CMV (i.e., bipolar characteristics). Hence the THD and leakage current in this topology is minimum. But, since the number of switches used in HERIC is more than that used in H5, the switching event in HERIC is high and hence it generates more harmonics than H5 [14].

*B. Methodology*

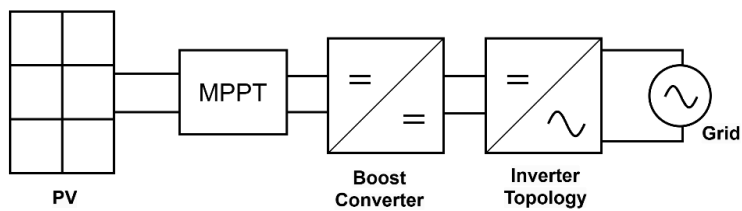

Fig. 12 Overall system block diagram

To validate the results of different topologies according to the existing literature, simulation models for each topology were developed and tested in MATLAB (Simulink) R2023a. The overall block diagram of the Simulink model is illustrated in Fig. 12.

A PV panel with a total power of 2.2kW is designed in Simulink. An MPPT block is used to track the maximum power point of the PV panel. Boost converter (DC-DC converter) precedes the MPPT block, and its main function is to regulate and adjust the output voltage of the PV panel, which is called DC-link voltage ($V_{pv}$). Since this overall design is based on a single phase, $V_{pv}$ is set constant at 400V. The switching signals for the inverter's switches are generated from the Hysteresis Band Current Controller (HBCC), which generates the switching signals to maintain desirable inverter output current. This whole system is done in a single-phase configuration, and a single-phase grid was also modeled in simulation.

## III. RESULTS

The results obtained from the simulation are discussed in this section. The first part of this section discusses the nature of CMV, DMV, and leakage current while the second part discusses the nature of THD. The circuit parameters used for the simulation are tabulated in Table 5.

Table 5: Circuit parameters

| **Parameter** | **Symbol** | **Value** |
|---|---|---|
| Input capacitor | $C_{DC}$ | 3000 µF |
| Filter inductors | $L_1, L_2$ | 4 mH |
| Parasitic capacitor | $C_{pv}$ | 24 nF |
| Input voltage | $V_{pv}$ | 400 V |

## A. CMV, DMV, and Leakage current results

In this part the nature of CMV, DMV, and the leakage current of all topologies are analyzed and are compared.

1) **H4 Unipolar Topology:** The waveforms of CMV, DMV, and leakage current for this topology are illustrated in the Fig. 13.

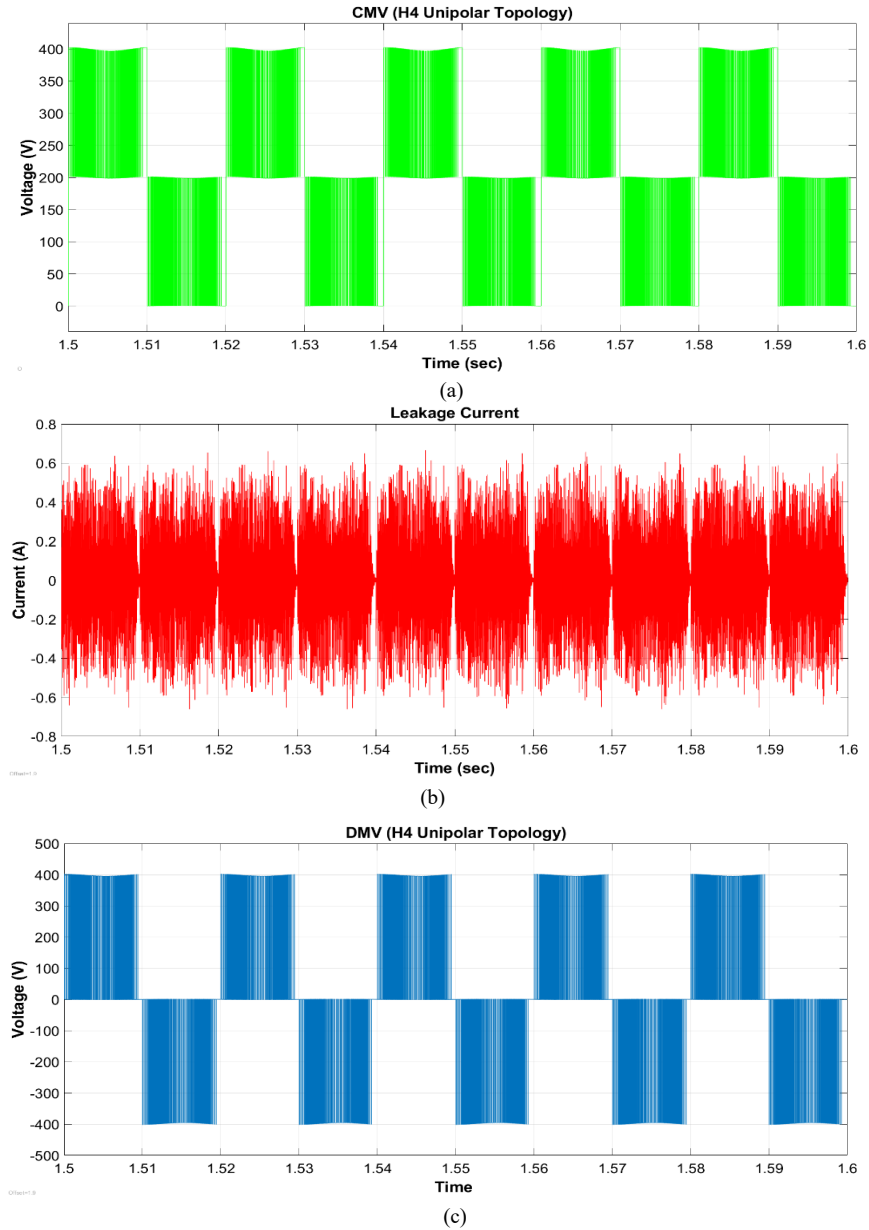

Fig. 13. (a) CMV, (b) leakage current, (c) DMV
H4 bipolar

From Fig. 13a, it can be seen that CMV of H4 topology with unipolar modulation is not constant and swings between three values $V_{PV}$, $V_{PV}/2$, and 0 i.e., 400V, 200V, and 0V (since $V_{PV}$ = 400V). Fig. 13c shows the waveform of DMV. DMV takes three distinct levels during one cycle complete cycle; these results comply with the result obtained in table 1. Since the CMV is not constant, there is a significant amount of leakage current flow in this topology. The waveform of the leakage current is shown in Fig. 13b, and its RMS value is 0.2697A.

2) *H4 Bipolar Topology:* The waveforms of CMV, DMV, and leakage current for this topology is illustrated in Fig. 14.

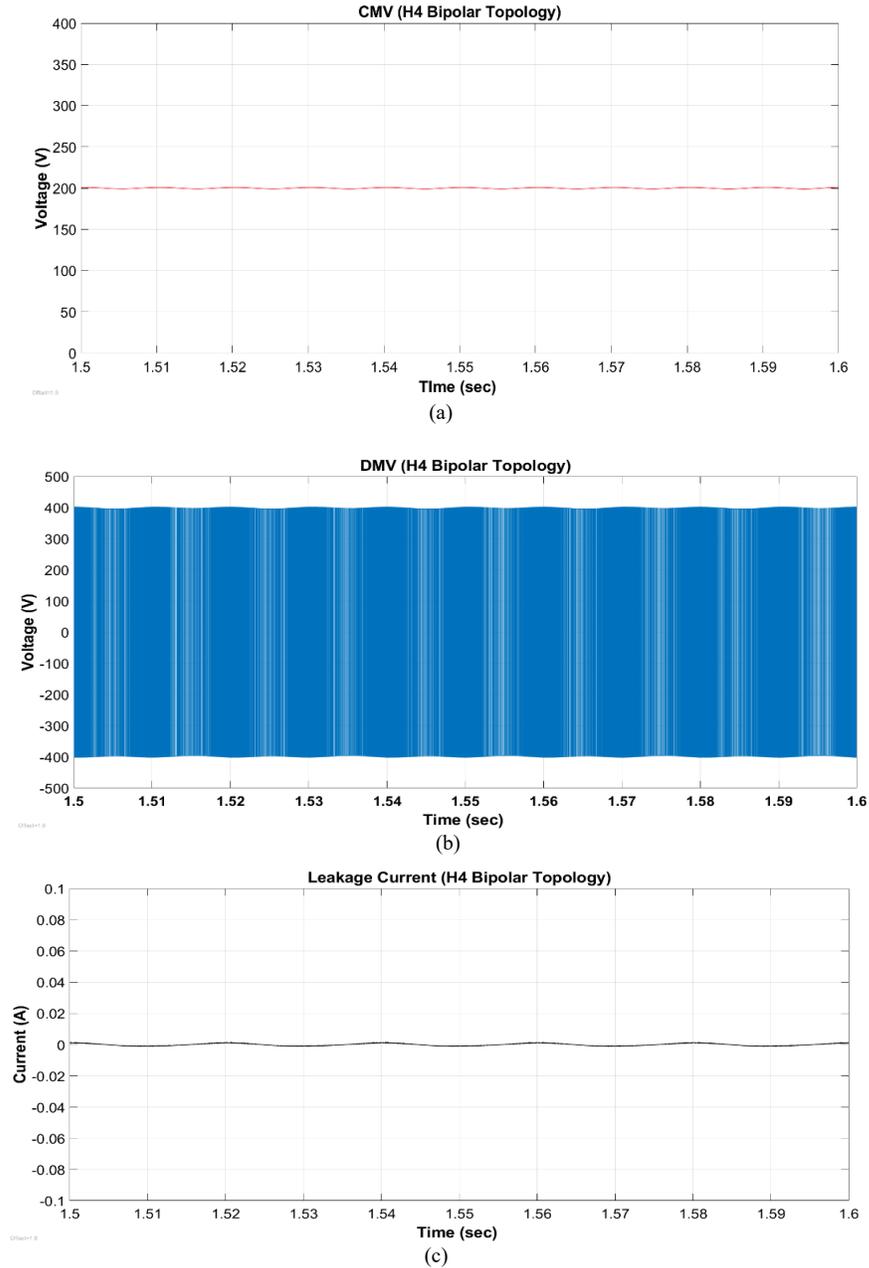

Fig. 14. (a) CMV, (b) DMV, (c) leakage current
H4 bipolar

Fig. 14a shows the waveform of CMV, which is constant and has a value of approximately 200V i.e., $V_{pv}/2$. Since the CMV is constant, the leakage current is approximately zero as shown in Fig. 14c. The RMS value of the leakage current obtained from the simulation is 0.6943mA. Fig. 14b shows the waveform of DMV and it can be seen that DMV has a two-level waveform i.e., it swings between $+V_{pv}$ to $-V_{pv}$ (+400V to -400V). This leads to higher THD and switching loss. All these simulation results comply with Table 2.

3) **H5 Topology:** The waveforms of CMV, DMV, and leakage current for this topology is illustrated in Fig. 15.

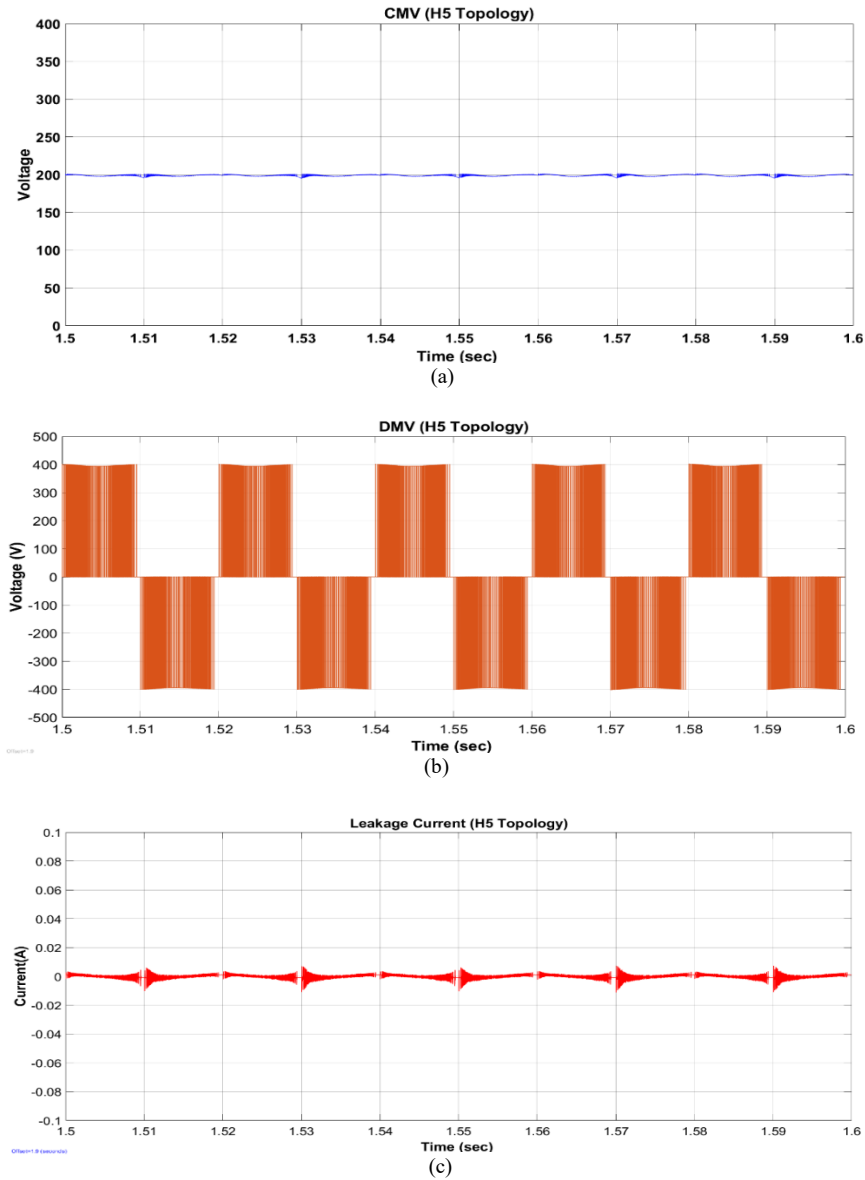

Fig. 15. (a) CMV, (b) DMV, (c) leakage current
H5

As depicted by Fig. 15a, the CMV in the H5 topology is approximately constant at 200V (i.e., $V_{pv}/2$). There are only minute spikes in CMV due to the use of two clamping diodes. Constant CMV ensures negligible leakage current, which can be seen in Fig. 15c. The RMS value of leakage current obtained from the simulation is 1.495mA. The DMV characteristics as shown in Fig. 15b has unipolar characteristics (i.e., it has three levels). So, the THD is also small in this topology, which makes it suitable for practical applications. All the simulation results comply with Table 3.

4) **HERIC Topology:** The waveforms of CMV, DMV, and leakage current for this topology is illustrated in Fig. 16.

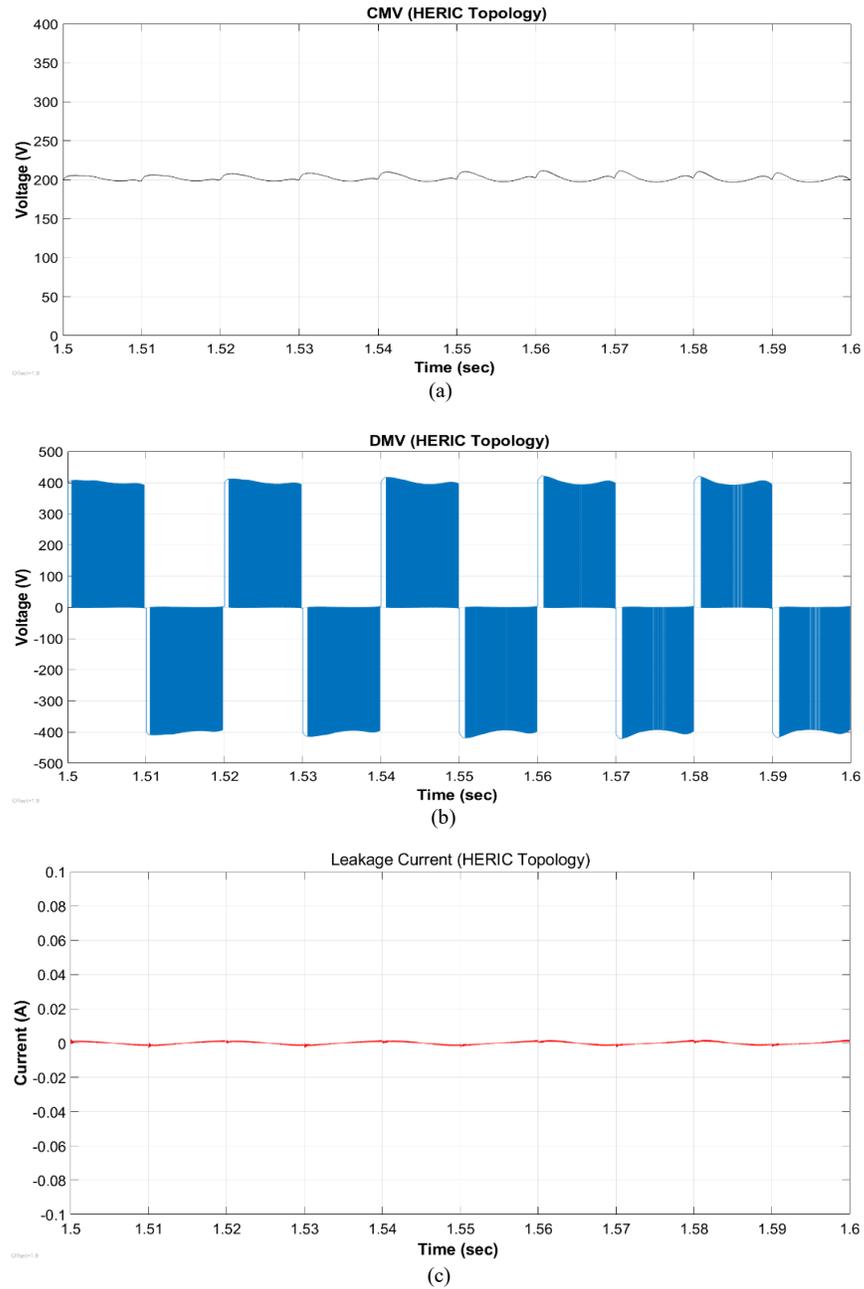

Fig. 16. (a) CMV, (b) DMV, (c) leakage current
HERIC

Fig. 16a confirms approximately constant CMV at the value of 200V. The DMV characteristics illustrated by Fig. 16b show the three-level voltage waveform. Hence, due to constant CMV, the leakage current is negligible, as verified by Fig. 16c, and due to the three-level voltage characteristics of DMV, the stress applied on the switches is small. But HERIC topology consists of one extra switch compared to H5 and it also has one additional switch switching at high frequency as shown in Fig. 10, so the THD distortion of this topology is higher than that of H5.

B. *Total Harmonic Distortion (THD) Analysis*

The THD of each topology is illustrated in Fig. 17. For the THD analysis, 40 cycles of grid current are examined. The fundamental frequency is taken at 50Hz and only up to third harmonics are analyzed.

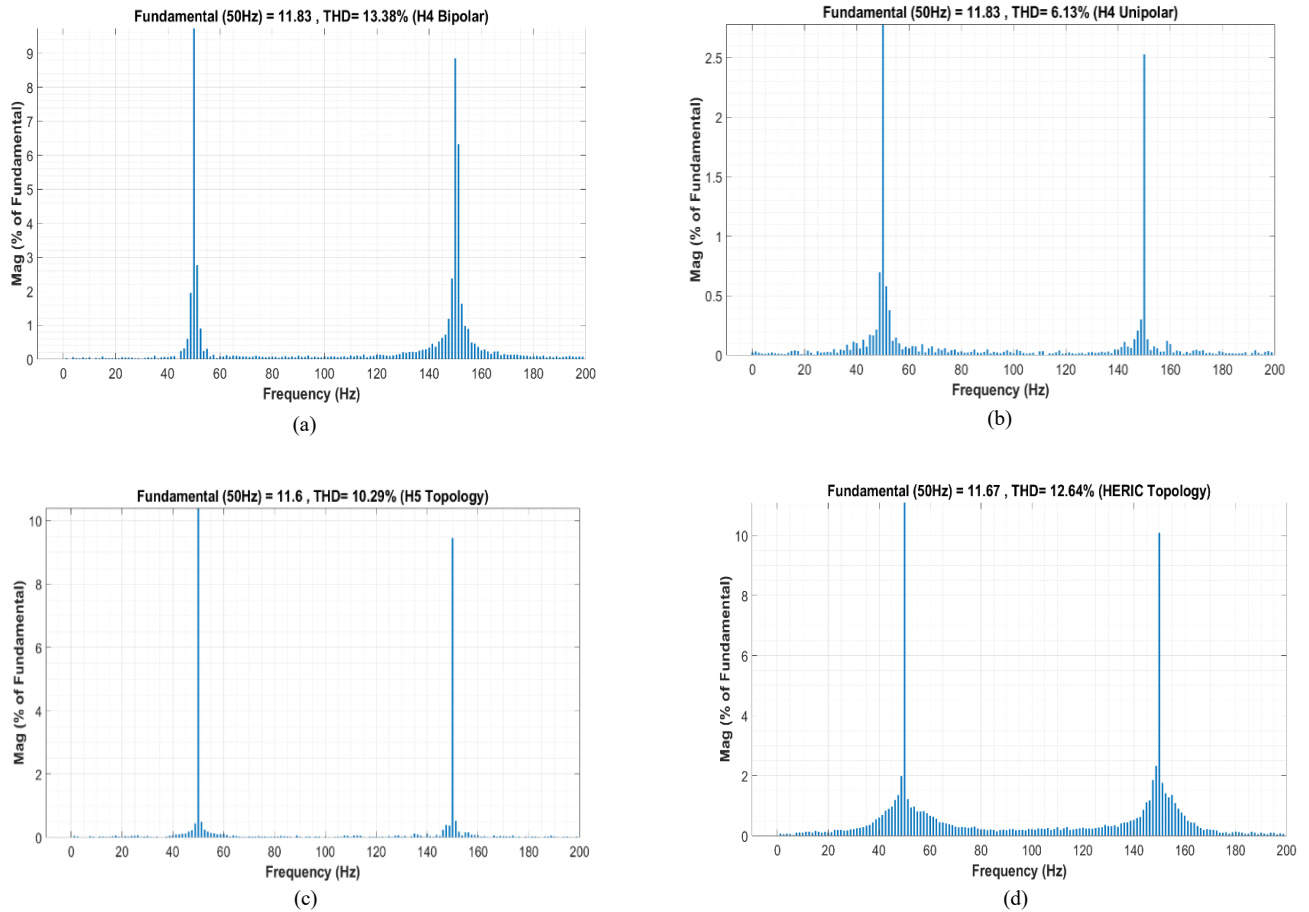

Fig. 17 Comparative FFT analysis of various inverter topologies: (a) H4 Bipolar, (b) H4 Unipolar, (c) H5 Topology, (d) HERIC Topology

The THD of the H4 topology with bipolar modulation is 13.38%, which is higher as compared with other topologies. It is because of the two-level voltage waveform of DMV generated by bipolar modulation as shown in Fig. 14b. On the other hand, the THD of H4 topology with unipolar modulation is 6.13%. This drastic change from 13.38% to 6.13% is due to the three-level voltage waveform of DMV in unipolar modulation illustrated by Fig. 13c. The THD of H5 topology is 10.29%. It is less than that of the H4 bipolar topology since in the former the DMV has a three-level voltage waveform. However, it is worth noting that the THD of the H5 topology is greater than that of the H4 unipolar topology. This is because in H5 topology one extra switch is used which contributes to increased harmonic distortion. Finally, the THD of HERIC topology is 12.64%, which is higher than that of H5 and H4 unipolar topology. It is because of two additional switches.

Hence, the H4 topology with unipolar modulation has the most favorable DM characteristics than other topologies. However, its CMV characteristics is not suitable for grid-connected transformer less inverters since a large leakage current circulates in the inverter circuit. H4 topology with bipolar modulation has the least magnitude of leakage current. However, the THD in this technique is very high due to the two-level DMV. Hence, the alternative topologies to maintain constant CMV and low THD are H5 and HERIC topologies. Both of these topologies maintain constant CMV hence minimizing the leakage current magnitude and the THD generated by both of these topologies is lower than that of the H4 bipolar topology. Table 6 summarizes the results obtained from the simulation for all topologies.

Table 6 Comparison table between H4 Unipolar, H4 Bipolar, H5, and HERIC TLI Topologies

| TOPOLOGY | $V_{DC}$ (V) | CMV | DMV | LEAKAGE CURRENT ($mA$) | NUMBER OF SWITCHES | THD |
|---|---|---|---|---|---|---|
| H4 UNIPOLAR | 400 | NOT CONSTANT | THREE-LEVEL | 296.7 | 4 | 6.13 |
| H4 BIPOLAR | 400 | CONSTANT | TWO-LEVEL | 0.6943 | 4 | 11.83 |
| H5 | 400 | CONSTANT | THREE-LEVEL | 1.495 | 5 | 10.29 |
| HERIC | 400 | CONSTANT | THREE-LEVEL | 0.78 | 6 | 12.64 |

## IV. CONCLUSION

This study presents a comparative analysis of transformer-less grid-connected PV inverter topologies such as H4, H5, and HERIC. The four main aspects for comparison are CMV, DMV, leakage current, and THD. The result demonstrates that although H4 topology with unipolar modulation produces the least harmonics distortion due to favorable DMV characteristics, it suffers from huge leakage current circulation due to the fluctuating nature of CMV. On the other hand, the H4 topology with bipolar modulation can generate constant CMV thereby reducing the leakage current but due to its two-level DMV, it generates significant harmonic distortion.

The H5 topology emerges as an optimized solution by introducing an additional switch at the positive DC bus and clamping diodes, effectively combining the unipolar modulation's DMV characteristics and the bipolar modulation's CMV characteristics. It reduces the leakage current and also maintains low harmonic distortion. Similar to the H5 topology, HERIC topology also maintains the CMV constant, but due to one extra switch switching at high frequency, it generates more harmonics distortion as compared to the H5 topology.

Overall, the H5 topology strikes the best balance between low leakage current, reduced THD, and operational efficiency, making it highly suitable for modern grid-connected PV systems.

## ACKNOWLEDGMENT

the authors would like to thank all individuals and organizations who supported this research. special thanks go to Professor Dr. Indraman Tamrakar for his invaluable guidance, insightful suggestions, and continuous support throughout this research. His expertise and mentorship were instrumental in the successful completion of this study.